\documentclass[prd,floats,preprint,superscriptaddress,eqsecnum,floatfix,nofootinbib,12pt]{revtex4}
\pdfoutput=1
\usepackage{amsmath}
\usepackage{graphics, graphicx}
\usepackage{epsfig}
\usepackage{ulem}
\usepackage{color}

\renewcommand{\Im}{{\rm Im}}
\renewcommand{\Re}{{\rm Re}}
\def\bh{\hat a}

\def\Nh{\hat N}
\def\Sh{\check S}

\def\Ih{\check I}

\def\1p{{(1p)}}
\def\be{\begin{equation}}
\def\ee{\end{equation}}
\def\beq{\begin{eqnarray}}
\def\eeq{\end{eqnarray}}
\def\p0{\phi_0}
\def\z0{\zeta_0}

\def\cV{{\cal V}}

\def\3G{^3{\cal G}}

\def\vx{{\vec x}}
\def\lm{\lambda_{-}}
\def\lp{\lambda_{+}}
\def\th{{\tilde h}}

\def\hij{h_{ij}(\vx)}
\def\hijt{{\tilde h}_{ij}(\vx)}
\def\chix{\chi(\vx)}
\def\Nh{{\hat N}}
\def\Ic{{\cal I}}
\def\bh{{\hat b}}

\def\Ich{{\hat{\cal I}}}
\def\ol2{\frac{1}{\ell^2}}

\def\cor{}
\def\ca{}
\def\cb{}
\def\cd{}
\def\ce{}
\def\crd{}
\def\cf{}
\def\cre{}
\def\cq{}
\def\df{}

\def\gf{}
\def\hf{}

\newcommand{\ttle}[1]{{\it #1}}

\begin{document}

\vspace{1cm}

\title{Accelerated Expansion from Negative $\Lambda$}

\author{James B.  Hartle}
\affiliation{Department of Physics, University of California, Santa Barbara,  93106, USA}
\author{S.W. Hawking}
\affiliation{DAMTP, CMS, Wilberforce Road, CB3 0WA Cambridge, UK}
\author{Thomas Hertog}
\affiliation{Institute for Theoretical Physics, KU Leuven, 3001 Leuven, Belgium {\it and}\\
International Solvay Institutes, Boulevard du Triomphe, ULB, 1050 Brussels, Belgium}

\bibliographystyle{unsrt}


\begin{abstract}
Wave functions specifying a quantum state of the universe must satisfy the constraints of general relativity, in particular  the Wheeler-DeWitt equation (WDWE). We show for a wide class of models with non-zero cosmological constant that solutions of the WDWE exhibit a universal semiclassical asymptotic structure for large spatial volumes. A consequence of this asymptotic structure is that a wave function in a gravitational theory with a negative cosmological constant can predict an ensemble of asymptotically classical histories which expand with a positive effective cosmological constant. This {\cd raises the possibility} that even fundamental theories with a negative cosmological constant can be consistent with our {\ce low-energy} observations of a classical, accelerating universe. 
We illustrate this general framework with the specific example of the no-boundary wave function in its holographic form. The implications of these results for model building in string cosmology are discussed.

\end{abstract}


\pacs{98.80.Qc, 98.80.Bp, 98.80.Cq, 04.60.-m CHECK PACS ADS}

\maketitle

\tableofcontents

\section{Introduction}
\label{intro}

The observed classical expansion of our universe is accelerating at a rate consistent with a positive cosmological constant of order $\Lambda \sim 10^{-123}$ in Planck units \cite{WMAP}. Does this tell us even the sign of the cosmological constant in the fundamental theory?  We present evidence that the answer to this question can be `no' and that even theories with a negative cosmological constant can predict accelerating classical histories. The underlying reason is that in quantum gravity the theory specifies a wave function of the universe, from which classical evolution emerges only in certain {\gf regions of superspace}. At the level of the wave function there is a close connection between asymptotic Lorentzian de Sitter (dS) spaces and Euclidean anti-deSitter (AdS) spaces\footnote{See \cite{Bala02,Maldacena03,Maldacena11,Harlow11,Anninos11,HH11} for earlier explorations of this connection.}.  We have argued that this AdS/de Sitter connection is more profound than -- and generally different from -- a continuation between solutions of one theory to solutions of a different theory. In particular we showed \cite{HH11} that  such a connection is manifest in the semiclassical approximation to the no-boundary quantum state  for one given dynamical theory, which can be taken to be a consistent truncation of AdS supergravity. In this paper we will show that this connection holds more generally for any state whose wave function satisfies the constraints of general relativity\footnote{A summary of these results is in \cite{HHH12b}.}.

It is common in cosmology to assume a classical background that solves the classical dynamical equations of the underlying theory. Fluctuations about that background are treated quantum mechanically. Under those assumptions the value of the cosmological constant governing the accelerated expansion has the same sign as that in the input theory\footnote{Sometimes the parameters of the background are analytically continued to give other backgrounds, but that does not address the question of what the possible observed backgrounds are.}.
However, in quantum cosmology {\it both} the backgrounds and the fluctuations are treated quantum mechanically. Then a non-trivial connection between the observed cosmological parameters and those of the input theory is possible. This paper investigates this possibility. 

We work largely with simple dynamical models in four dimensions consisting of spatially closed four-geometries with metric $g_{\alpha\beta}(x)$ coupled to a single scalar field $\phi(x)$. {\cd A specific model to which our analysis applies is the consistent truncation of the low energy limit of M theory compactified on $S^7$ which involves only AdS gravity and a single scalar field with a negative potential (see e.g. \cite{Hertog05}).} In this context, a quantum state of the universe is specified by a wave function $\Psi$ on the superspace of  three-geometries ($h_{ij}(\vx)$)  and matter field configurations ($\chi(\vx)$) on a closed spacelike  three-surface $\Sigma$. Schematically we write $\Psi=\Psi[h,\chi]$. 

In some regions of superspace the quantum state may be approximated to leading order in $\hbar$  by a sum of terms of semiclassical (WKB) form 
\begin{equation}
\Psi[h,\chi] \propto  \exp(-I[ h,\chi]/\hbar)\equiv \exp(-I_R[h,\chi] +i S[h,\chi])/\hbar) . 
\label{semiclass}
\end{equation}
Here $I$ is a complex ``action" functional and $I_R$ and $-S$ are its real and imaginary parts. 
{\cd Classical cosmological evolution emerges from the quantum state} in regions of superspace where $S$ varies rapidly compared to $I_R$ (as measured by quantitative classicality conditions \cite{HHH08}). This is analogous to the prediction of the classical behavior of a particle in a WKB state in non-relativistic quantum mechanics. When there are regions of superspace where the classicality conditions hold a quantum state predicts an {\it ensemble} of spatially closed classical Lorentzian cosmological histories that are the integral curves of $S$ in superspace.  That means that the histories are determined by  the Hamilton-Jacobi relations relating the momenta proportional to time derivatives of metric and field to the superspace derivatives of $S$.
Their relative probabilities are also determined by the wave function. To leading order in $\hbar$ they are proportional to $\exp(-2I_R[h,\chi]\hbar)$, which is constant along the individual integral curves\footnote{The analogy to WKB in non-relativistic quantum mechanics makes this prescription for classical predictions in quantum cosmology plausible. It can be derived to a certain extend from the generalized decoherent histories quantum mechanics of cosmological histories \cite{Har95c}.}. 
 
The case of the no-boundary wave function (NBWF) \cite{HH83} is a familiar illustration of how semiclassical approximations of the form  \eqref{semiclass} can arise. In the NBWF the functional $I$ is the action of a complex saddle point of the underlying Euclidean action that is regular on a four-disk and matches $(\hij, \chix)$ on its boundary. A unified structure for different saddle points is provided by the complex solutions of the Einstein equations that are a necessary condition for an extremum of the action. These complex solutions have domains where the metric and field are real --- real domains. These real domains are candidates for predicted classical histories provided that the classicality condition that $S$ varies rapidly in comparison with $I_R$ is satisfied. 
 
However, we show in this paper that the complex structure defined by NBWF saddle points is but a special case of a more general complex structure defined by the Wheeler-DeWitt equation (WDWE) and therefore common to all cosmological wave functions.  In particular we find that, when the cosmological constant is non-zero, the WDWE implies that the semiclassical approximation \eqref{semiclass} is valid asymptotically for large spatial volumes. Further, the leading terms in the asymptotic expansion of the action $I$ are common to all solutions. That is, they are fully determined by the final configuration of metric and field on $\Sigma$ and independent of the boundary conditions at small scale factor implied by a specific choice of quantum state. This universal structure is the analog for wave functions of the {\gf leading universal terms in the} Fefferman-Graham \cite{Fefferman85} {\cf and Starobinsky \cite{Starobinsky83}} asymptotic expansions of solutions to the Einstein equations. Indeed it is essentially the same thing since there is a close connection between actions and solutions of the equations of motion.  With this more general WDWE structure we can see that a number of results obtained in \cite{HH11} for the NBWF are in fact properties of any cosmological wave function. 

We find that the semiclassical asymptotic wave function in a negative $\Lambda$ theory, evaluated on a boundary with spherical topology\footnote{{\cf For recent work on the wave function of the universe evaluated on boundaries with different topologies see e.g. \cite{Castro11}.}}, includes two classes of real domains which are either asymptotically Euclidean AdS in one signature or Lorentzian de Sitter in the opposite signature. It is natural to search both real domains for the predictions of classical histories in the ensemble by checking the classicality condition. This is suggested because they both emerge from the same complex structure and because there is no physical reason to prefer one signature over another. More importantly it is suggested by holography \cite{Maldacena98}, especially in the case of the NBWF. 
 
In its holographic form \cite{HH11}, the semiclassical no-boundary {\cor wave function} is a product of a factor involving the partition function of an {\cf AdS/CFT dual Euclidean} field theory on the conformal boundary and a universal factor that is fully determined by the argument of the wave function\footnote{{\cf For earlier discussions of Euclidean AdS/CFT viewed as a statement about the wave function of the universe see e.g. \cite{Maldacena03,Horowitz04}.}}. While the former factor governs the relative probabilities of different histories, the validity of the semiclassical approximation follows from the universal factor alone. That is because this provides the leading behavior of the wave function in the large volume regime. {\gf At this level}, the dual field theory is insensitive to the signature of the metric because it is conformally invariant. This supports treating both real domains equally. 

We find that the classicality condition can be satisfied for the de Sitter real domain. Thus negative $\Lambda$ theories predict accelerated expansion for a wide class of wave functions as a consequence of the WDWE. {\cf The relative probabilities of different asymptotic de Sitter histories (including their perturbations) are given by the dual partition function or, via AdS/CFT, by the AdS regime of the theory. The wave function of the universe thus provides a framework in which the holographic calculations of CMB correlators \cite{Fadden10, Pimentel11} can be put on firm footing and generalized to models where the AdS/de Sitter connection differs from a simple analytic continuation.}

In Sections \ref{mss} and \ref{general} we first derive the universal asymptotic behavior of the wave function from the WDWE. This yields a universal asymptotic form of solutions to the Einstein equations. The asymptotic expansions suffice to show that quantum states obeying the WDWE in a negative $\Lambda$ theory imply an ensemble of classical histories which expand, driven by an `effective' positive {\cor cosmological constant equal to}  $-\Lambda$.

While the presence of expanding histories is general, the probabilities for the individual classical histories depend on the specific wave function. We illustrate how to calculate these probabilities in Section \ref{NBWF} for the no-boundary wave function, defined holographically for negative $\Lambda$ in terms of the partition functions of  (Euclidean) AdS/CFT duals \cite{HH11}. We find that in the large volume regime, the classical expanding histories give the dominant contribution to the NBWF.

\section{Asymptotic Semiclassical Structure of the Wheeler-DeWitt Equation in Minisuperspace}
\label{mss}

We begin with the simplest minisuperspace model to illustrate what an asymptotic structure is, and to give an explicit derivation of its form. In this model there is a negative cosmological constant $\Lambda$ but  no matter field $\phi$.  The geometries are restricted to the class of closed, homogeneous, isotropic cosmological models. The spatial geometries on a three-sphere are then characterized by a scale factor $b$. Cosmological wave functions are functions of $b$, $\Psi=\Psi(b)$. {The generalization to include a scalar field in this homogeneous, isotropic minisuperspace models is discussed in \cite{HHH12b}}.

\subsection{Geometry and Action}
\label{geomandaction}

We consider complex metrics describing four-geometries foliated by compact homogeneous and isotropic spatial slices
written in the form
\be
ds^2 =N^2(\lambda)d\lambda^2 +a^2(\lambda)  d\Omega^2_3
\label{4metric}
\ee
where $(\lambda, x^i)$ are four real coordinates on a manifold $M={\bf R}\times S^3$ and $d\Omega^2_3$ is the line element on the unit, round, three-sphere. 

Our analysis \cite{HH11} of the semiclassical approximation to the no-boundary state shows that it is specified by  complex solutions of the Einstein equations with asymptotically real domains. {\ca Different kinds of real domains can be represented by the metric \eqref{4metric}}. Real $N$ and $a$ represent positive signature\footnote{In this paper positive signature means that the signature of the three-metric is $(+,+,+)$; negative signature means $(-,-,-)$.}, Euclidean four-geometries; imaginary $N$ and real $a$ represent positive signature Lorentzian four-geometries. Real $N$ and imaginary $a$  provide a negative signature representation of Lorentian four-geometries; imaginary $N$ and $a$ provide a negative signature representation of Euclidean four-geometries.  By considering complex metrics of the form \eqref{4metric} we attain a unified description of all these cases. 
The action functional for this minisuperspace model {\cor can be taken to be}
\begin{equation}
I[N(\lambda),a(\lambda)] = \eta \int d\lambda N  \left[\frac{1}{2} G(a) \left(\frac{a'}{N}\right)^2  +\cV (a) \right] .
\label{mini-act}
\end{equation}
Here, $\eta\equiv3\pi/2$, $f'=df/d\lambda$ and
\be
G(a) \equiv -a, \quad \cV(a)\equiv - \frac{1}{2}(a +\frac{1}{\ell^2}a^3) 
\label{defGV}
\ee
with $1/{\ell^2}\equiv -\Lambda/3$ so that  $\ell$ is the usual AdS radius.  

\subsection{Wheeler-DeWitt Equation}
\label{WDWE}

Any wave function must satisfy the operator forms of the four-constraints of general relativity. The three momentum constraints are satisfied automatically as a consequence of the symmetries of this very simple model.  The operator form of the remaining Hamiltonian constraint is the Wheeler-DeWitt equation.   To derive  its form it is convenient to start with imaginary $N$ ($N=i{\hat N}$, real $\Nh$). Then \eqref{mini-act} summarizes the Einstein equations for Lorentzian histories and the familiar definitions of momenta apply.

Variation of the action with respect to $\Nh$ gives the classical equation
\be
\label{constraint}
\left(\frac{a'}{\Nh}\right)^2 +1 + \frac{1}{\ell^2} a^2 =0 .
\ee
Expressing this in terms of the momentum $p_a$ conjugate to $a$ gives
\be
\label{constraint}
\left(\frac{p_a}{\eta a}\right)^2 +1 + \frac{1}{\ell^2} a^2 =0 .
\ee
The operator form of this results from replacing $p_a$  by $-i\hbar(\partial/\partial a)$.
 Choosing the simplest operator ordering\footnote{As we will see in Section \ref{asymsemiclass} the asymptotic semiclassical structure depends only on the classical Hamilton-Jacobi equation and is independent of the operator ordering adopted.}  gives the WDWE
\be
\left(-\frac{\hbar^2}{\eta^2} \frac{d^2}{d b^2} + b^2 +\ol2 b^4\right)\Psi(b) = 0  . 
\label{mini-wdw}
\ee
Here, we have replaced $a$ by $b$ to emphasize that this is a relation dealing with three-metrics. 
We now investigate the solutions of this equation for large $b$.

\subsection{Asymptotic Semiclassicality}
\label{asymsemiclass}
For this simple model,  the WDWE \eqref{mini-wdw} has the same form as a zero energy, time-independent Schr\"odinger equation with a potential that diverges like $b^4$ at large $b$. A semiclassical approximation will hold there because this is a regime where the familiar WKB approximation applies. To exhibit it explicitly we write the wave function in the form
\be 
\Psi(b)\equiv \exp[-\Ic (b) / \hbar] \equiv \exp[-\eta\Ich(b)/\hbar] \  .
\label{defI}
\ee
The notation $\Ic$ is intended to suggest a semiclassical approximation, but at this point  we mean \eqref{defI} to be a {\it definition} of $\Ic(b)$ (and of  $\Ich(b)$)  which in general will be $\hbar$ dependent. In terms of $\Ich$, the WDWE \eqref{mini-wdw} becomes
\be
\frac{\hbar}{\eta}\frac{d^2\Ich}{db^2} - \left(\frac{d\Ich}{d b}\right)^2 + b^2 +\frac{1}{\ell^2}b^4 =0 \ .
\label{wdwI}
\ee
In regimes of $b$ where the first term is negligible, the WDWE reduces to the classical Hamilton-Jacobi equation with $\Ich\equiv \Ih$
\be
- \left(\frac{d \Ih}{d b}\right)^2 + b^2 +\ol2 b^4 =0
\label{mini-HJ}
\ee
which is independent of $\hbar$. Substituting a solution of  \eqref{mini-HJ} into \eqref{defI} gives a leading order in $\hbar$ semiclassical approximation to the wave function.

The condition for the first term in \eqref{mini-HJ} to be negligible and a semiclassical approximation to $\Psi$ to be valid\footnote{This condition for a semiclassical approximation to the wave function is weaker than the classicality condition for classical histories discussed in the Introduction which requires in addition that the imaginary part of $I$ vary more rapidly than the real part \cite{HHH08}.} 
{\cor is the semiclassicality condition}
\be
\frac{\hbar}{\eta}\left| \frac{d^2\Ich}{db^2}\right| \ll \left(\frac{d \Ich}{d b}\right)^2 \  .
\label{sccond}
\ee
We can see {\cor self consistently} how this condition is satisfied for large $b$ by first assuming it holds, then solving the Hamilton-Jacobi equation \eqref{mini-HJ}, and then checking that the solution satisfies the condition.

Asymptotically in $b$, {\df solutions} to the Hamilton-Jacobi equation \eqref{mini-HJ} have the expansion 
\be
\Ih(b)=\pm\left[\frac{1}{3\ell} b^3 + \frac{\ell}{2} b + c_{0}^{\pm} +{\cal O}\left(\frac{1}{b}\right)\right]
\label{asymptact}
\ee
where $c_{0}$ is a constant not fixed by \eqref{mini-HJ}. {\df This means it can also depend on the $\pm$ sign multiplying the series in \eqref{asymptact} as the superscript indicates.} The coefficient of the $b^2$ term is zero. It is then straightforward  to check that the semiclassicality condition \eqref{sccond} is satisfied for the powers $b^3$, $b^2$ and $b$ in \eqref{asymptact}. These first three powers constitute the universal asymptotic semiclassical structure of wave functions implied by the quantum implementation of the constraints of general relativity.

It is important to stress that this asymptotic semiclassical behavior holds for {\it any} wave function of the universe in this minisuperspace model. The coefficients of the first three powers of the asymptotic expansion \eqref{asymptact} are independent of any boundary condition that might single out a particular wave function. (One at small $b$ for example.) These coefficients are universal. They are the analog for wave functions of the Fefferman-Graham universal asymptotic structure of solutions to the Einstein equation \cite{Fefferman85,Skenderis02}. By contrast, the coefficient $c_{0}$  will depend on the specific wave function. 

\subsection{Asymptotic Structure of Solutions to the Einstein Equations}
\label{asymp-einstineq}
There is close connection between solutions of the Hamilton-Jacobi equation \eqref{mini-HJ} and solutions of the equations of motion following from the action \eqref{mini-act}.  Solutions to the equations of motion can be constructed from solutions to the Hamilton-Jacobi equation. Conversely,  substituting solutions of the equations of motion into \eqref{mini-act} gives an action that satisfies the Hamilton-Jacobi equation. The universal semiclassical structure we deduced from the Hamilton-Jacobi equation \eqref{mini-HJ} at large $b$ must therefore have an equivalent expression as an asymptotic solution to the Einstein equations of motion. We will now construct that correspondence explicitly for this minisuperspace model.

The relation between momenta and the gradient of the action $p=\nabla I$ provide the equations for finding solutions of the equations of motion from solutions of the Hamilton-Jacobi equation. For our simple minisuperspace model the relevant connection is
\be
p_b \equiv\eta \frac{G(b)}{N}\frac{db}{d\lambda}=\frac{dI}{d b} = \eta \frac{d \Ih}{d b} .
\label{mini-eqmotion}
\ee
Choosing $N$ real gives Euclidean solutions; chosing it imaginary gives Lorentzian ones. It is convenient to provide a unified description of these cases by introducing a parameter $\tau$ defined by  $d\tau=N(\lambda)d\lambda$ and considering solutions for complex $\tau=x+iy$.   Equation \eqref{mini-eqmotion} then takes the simple form
\be
\frac{db}{d\tau}=-\frac{1}{b}\frac{d\Ih}{db} .
\label{eqmotion}
\ee
Solutions representing real geometries are curves in the complex $\tau$-plane along which $b^2$ is real. 

Inserting the asymptotic expansion for the action \eqref{asymptact}  in \eqref{eqmotion}  gives an equation that can be systematically solved for large $b$. The variable $u\equiv \exp(-\tau/\ell)$  is convenient for exhibiting the asymptotic solution at large $x$ which is an expansion about small $u$ of the form
\be
b(u)=\ell\frac{c}{u}\left(1-\frac{u^2}{4c^2} +\cdots \right) .
\label{asympt-b}
\ee
 The {\cor complex } constant $c$ is undetermined. Once it is fixed, the coefficients of the next two powers of $u$ are determined. This is the universal asymptotic structure of the equations of motion.

It was not necessary to first construct the action to find the universal equation of motion structure. That emerges directly from an analysis of the asymptotic solutions of the Einstein equation. That asymptotic behavior has been worked out  in complete generality (e.g. \cite{Fefferman85,Starobinsky83,Skenderis02}).  However, to make use of this classical structure in a quantum mechanical context requires that it be connected to quantum amplitudes. That connection is provided by the  equivalent asymptotic semiclassical structure of solutions of the WDW equation together with the essentially quantum condition \eqref{sccond} that determines when this structure holds and how it is limited.

\subsection{Predictions for Classical Histories}
The asymptotic semiclassical structure of the WDWE can be used to investigate the predictions of any wave function for asymptotically classical histories. We recall from the discussion in the Introduction (eq. \eqref{semiclass} in particular) that classical histories are predicted in regions of the superspace of real three-geometries where the imaginary part of the action $-S$ varies rapidly compared with the real part $I_R(b)$. The histories are then the integral curves of $S$ and their relative probabilities are proportional to $\exp(-2I_R/\hbar)$ to leading order in $\hbar$ (tree level). Real three metrics correspond to real $b$ or imaginary $b=i\bh$. As discussed in the Introduction,  we search for classical histories in both {\ca of these real domains}. 

When $b$ is real the leading imaginary term in the  asymptotic action \eqref{asymptact} is $\Im(c_{0})$ which does not vary rapidly with respect to the real part. Therefore no classical geometries are predicted in this {\ca real domain}. 

When $b$ is purely imaginary ($b=i\bh$ with $\bh$ real) the WDWE \eqref{mini-HJ} implies the following expansion of the action,
\be
\Ih(\bh) = \pm i \left[ \frac{1}{3\ell}\bh^3 - \frac{\ell}{2}\bh +i \hat c_{0}^{\pm} + {\cal O}\left(\frac{1}{\bh}\right)\right] \ .
\label{asymptact-neg}
\ee
where $\hat c_{0}^{\pm}$ is a (generally complex) constant that is not determined by \eqref{asymptact}.
The leading real and imaginary parts of the asymptotic action in this domain are 
\begin{subequations}
\label{real-imag}
\begin{align}
\Sh(\bh)=&   \pm  \left[ \frac{1}{3\ell}\bh^3 - \frac{\ell}{2}\bh +\Im(\hat c_{0}^{\pm}) \right]  \label{imag} \\ 
\Ih_R(\bh)=& \mp  \Re(\hat c_{0}^{\pm})
\label{real}
\end{align}
\end{subequations}
The classicality condition that $S$ vary rapidly with $b$ compared to $I_R$ is easily satisfied for large $\bh$. The variation in  $S$ is large and $I_R$ doesn't vary at all. 

The integral curves of $S$ are the solutions of 
\be
\frac{d\bh}{dt} = \frac{1}{\bh}\frac{d\Sh}{d\bh}
\label{intcurveq}
\ee
(cf. \eqref{mini-eqmotion} with N=-1 for convenience). The asymptotic solution for large $b$ can be put in the form
\be
\bh(t) = \frac{\ell}{2}[e^{(t-t_*)/\ell} + e^{-(t-t_*)/\ell}] + {\cal O}(e^{-2t/\ell})
\label{deSitter}
\ee
by appropriate definition of the integration constant $t_*$.  Thus we predict the classical Lorentzian history described by the metric
\be
ds^2=dt^2 - \bh^2(t)d\Omega_3^2
\label{negdS}
\ee
This describes an asymptotic, Lorentzian deSitter expansion controlled by a positive cosmological constant $3/\ell^2$. 

The relative probability assigned to this history by our prescription for classical prediction is  proportional to $\exp[-2\eta Re(c_{0})/\hbar]$. This does not mean much for one history. The normalized probability is $1$. We will get more non-trivial examples when matter is included in Section \ref{general}.

We therefore reach the striking conclusion in a minisuperspace model that, for gravitational theories with negative cosmological constant, all wave functions predict one asymptotic classical history with a deSitter expansion governed by the magnitude of the cosmological constant. In Section \ref{general} we will show that this conclusion can hold much more generally than in the minisuperspace models considered here.

\subsection{Signature Neutrality}
\label{signeutrality}

As already mentioned, the overall sign of the four-metric is not a physically measurable quantity.   We can observe that the universe is expanding but not the overall sign  of the metric representing that expansion. That is because measurable properties of geometry are ratios of distances, not the distances  themselves. Physical theories, including those of the wave function of the universe, must be signature neutral --- not preferring one overall sign to another.

Yet the metrics \eqref{negdS} describing classical Lorentzian accelerating universes were predicted with a specific  signature  --- negative. 
No classical univeses with positive signature were predicted. The origin of this signature asymmetry can be traced to the positive signature conventions assumed in writing down the action in the form \eqref{mini-act}. Had we started with the opposite convention for the action the overall sign in \eqref{negdS} would have been reversed. The theory as formulated here is therefore  signature neutral but not signature invariant.

\section{Beyond Minisuperspace}
\label{general}

\subsection{Geometry and Action}
\label{geomact-gen}
This section generalizes the results of the simple minisuperspace model  to general metrics.  We consider the class of four-dimensional models with Einstein gravity  coupled to a single scalar field with an everywhere negative potential $V(\phi)$. {\df We take the mass $m^2$ to be negative but within the Breitenlohner-Freedman (BF) range $m^2_{BF} < m^2 <0$, where $m^2_{BF} = -9/(4\ell^2)$.} The metric $g(x)$ (short for $g_{\alpha\beta}(x^\gamma)$) and $\phi(x)$ are the histories of the 4-geometry and matter field. The Euclidean action $I[g(x),\phi(x)]$ is the sum of the Einstein-Hilbert action and the matter action
\be
\label{actionsum}
I[g(x),\phi(x)] = I_C[g(x)] + I_{\phi}[g(x),\phi(x)] .
\ee
Here, specifically, with a positive signature convention for the metic, 
\begin{subequations}
\label{action}
\begin{equation}
I_C[g] = -\frac{1}{16\pi}\int_M d^4 x (g)^{1/2}\left(R+\frac{6}{\ell^2}\right) -\frac{1}{8\pi}\int_{\partial M} d^3 x (h)^{1/2}K
\label{curvact}
\end{equation}
and 
\begin{equation}
I_{\phi}[g,\phi]=\frac{3}{8\pi} \int_M d^4x (g)^{1/2}[(\nabla\phi)^2 +V(\sqrt{4\pi/3}\phi)] \ .
\label{mattact}
\end{equation}
\end{subequations}
The normalization of the scalar field $\phi$ has been chosen to simplify subsequent equations and maintain consistency with our earlier papers. {\df It differs from the usual normalization by a factor $\sqrt{4\pi/3}$ \cite{HHH08}.} In a suitable gauge the metric can be written
\begin{equation}
ds^2=N^2(\lambda) d\lambda^2 +h_{ij}(\lambda,\vx) dx^i dx^j \ .
\label{eucmetric_hij}
\end{equation}
This generalizes \eqref{4metric}. 
The large volume asymptotic form of the general solutions of the Einstein equations has been worked out by Fefferman and Graham \cite{Fefferman85}. 
Introducing $d\tau = N(\lambda) d\lambda$  and defining $u\equiv \exp(-\tau/l)$, the metric expansion for small $u$ reads
\begin{subequations}
\label{expansions}
\be
\label{hijexpn}
h_{ij}(u,\vx)=\frac{c^2}{u^2}[\th^{(0)}_{ij}(\vx) +\th_{ij}^{(2)}(\vx) u^2 + \th_{ij}^{(-)}(\vx)u^{\lambda_{-}} +\th_{ij}^{(3)}(\vx)u^3 +\cdots] . 
\end{equation}
where $\th^{(0)}_{ij}(\vx)$ is  real and normalized to have unit volume thus determining the constant $c$. This generalizes \eqref{asympt-b}.
For the field one has
\be
\label{phiaexpn}
\phi(u,\vx) = u^{\lm}(\alpha(\vx) +\alpha_1(\vx) u + \cdots)  +  u^{\lp}(\beta(\vx) +\beta_1(\vx) u +\cdots) . 
\ee
\end{subequations}
where $\lambda_{\pm} = \frac{3}{2}(1\pm \sqrt{1+(2m/3)^2})$ with $m^2$ the (negative) scalar mass squared.

As with \eqref{asympt-b} the asymptotic solutions are locally determined from the asymptotic equations in terms of the `boundary values' $c^2 \th^{(0)}_{ij}$ and $\alpha$, up to the $u^3$ term in \eqref{hijexpn} and to order $u^{\lambda_{+}^{\ }}$ in \eqref{phiaexpn}.  {\gf Hence the leading behavior of asymptotic solutions is universal.} Beyond this order the interior dynamics and the relevant  boundary conditions  become important. 

The wave function is a functional of three-metics $\hij$ and field configurations $\chi(\vx)$ on a spacelike three surface.  It will prove convenient to separate an overall scale from the three metric by writing 
\be
\hij = b^2 \hijt
\label{bdef}
\ee 
and requiring $\hijt$ to have unit volume. Asymptotically in solutions $\hij$ coincides with $h^{(0)}(\vx)$ as \eqref{hijexpn} shows. 

Coordinates spanning superspace are then $(b,\hijt,\chi(\vx))$.  We denote these collectively by $q^A$, or when we want to separate $b$ from the others, by $(b,\theta^i(\vx))$. In the $c=G=1$ units used throughout, the coordinate $b$ has dimensions of length and the other coordinates $\theta^i$ are dimensionless. Dimensions of quantities are thus related to powers of $b$. 

In terms of these coordinates, the action $I$ defined by \eqref{actionsum} can be expressed as 
\begin{equation}
I = \int d\lambda d^3x  N   \left[\frac{1}{2} G_{AB}(q^A) \left(\frac{1}{N}\frac{dq^A}{d\lambda}\right) \left(\frac{1}{N}\frac{dq^B}{d\lambda}\right) +\cV (q^A) \right] .
\label{mini-eucact}
\end{equation}
Here, $G_{AB}(\vx)$ is the DeWitt supermetric on superspace and $\cV$ incorporates the cosmological constant and the matter potential $V$. All quantities under the integral sign are functions of $\vx$. Eq \eqref{mini-eucact}  is the generalization of \eqref{mini-act}. Variation with respect to $N$ gives the Hamiltonian constraint generalizing \eqref{constraint} and replacing momenta by operators in that gives the WDWE generalizing \eqref{mini-wdw}. 

\subsection{Semiclassicality}
\label{semiclass-gen}
{\gf The expansions \eqref{expansions} have been used in the context of AdS/CFT to obtain the leading behavior of the large volume asymptotic form of the action for real solutions of the Einstein equations (see e.g. \cite{Skenderis02}). The contributions to the action that grow as a function of the scale factor are universal and yield the so-called counterterms.} These can be written as\footnote{The coefficient of the scalar counterterm differs from \cite{Skenderis02} because we work with a rescaled scalar field variable (cf. eq \eqref{mattact}.} 
\be
\label{ct}
S_{ct} [h,\chi] =  \frac{1}{4\pi l}\int  d^3x \sqrt{h} + \frac{l}{16\pi} \int d^3x \sqrt{h} \ {{^3}R(h)} +\frac{3\lambda_{-}}{8\pi l} \int  d^3x \sqrt{h} \chi^2
+ \cdots
\ee
where the dots indicate higher derivative (gradient) terms. 

{\gf In field theory applications of AdS/CFT the counterterms \eqref{ct} are interpreted as UV divergences  in the field theory and subtracted  in order to regulate the volume divergences of the AdS action.} A regulation of the action by adding counterterms is not part of our framework \cite{HH11} but, {\it assuming a semiclassical approximation is valid}, the universal asymptotic form of the action obtained this way also defines the asymptotic solution of the WDWE as discussed in the preceding section. In particular the counterterms \eqref{ct} yield a series for the action $I$ in powers of $b$,
\be
I(b,\theta^i(\vx)] = b^3 c_3[\theta^i(\vx)] + b^2 c_2[\theta^i(\vx)] +  b c_1[\theta^i(\vx)]+ \cdots 
\label{actexpn}
\ee
The leading terms of the coefficient functionals  $c_n[\theta^i(\vx)]$ are determined by the boundary values $c^2 \th^{(0)}_{ij}$ and $\alpha$ of metric and field. This is the generalization of {\gf the first two terms} in \eqref{asymptact}. 

{\hf This expansion can be used to show that the generalization of the semiclassicality condition \eqref{sccond} is indeed satisfied. This is} 
\be
\label{sccond-gen}
\hbar G^{AB}(\vx)\frac{\delta^2I}{\delta q^A(\vx) \delta q^B(\vx)} \ll G^{AB}(\vx) \frac{\partial I}{\partial q^A(\vx)}\frac{\delta I}{\delta q^B(\vx)} \ .
\ee
Eq \eqref{actexpn} can be used to expand both  the right and left hand side of \eqref{sccond-gen} in powers of $b$. Because $b$ has dimensions of length and $\hbar$ has dimensions of length squared,  the expansion of the left hand side of \eqref{sccond-gen} must begin two powers of $b$ lower than that of the right hand side. That means that for very large $b$ the condition \eqref{sccond-gen} will be satisfied for the first three powers of $b$. 

Thus a universal, semiclassical asymptotic structure for solutions to the WDWE has been established self-consistently.  {\hf The asymptotic form of solutions to the Einstein equations \eqref{expansions} was assumed in order to derive the expansion of the action \eqref{actexpn} evaluated on these solutions.  That expansion satisfied the semiclassicality condition \eqref{sccond-gen} which in turn justified the use of the asymptotic expansions.} The asymptotic expansion of this action and the asymptotic expansion of the Einstein equations are equivalent expressions of the same structure. 
  
\subsection{Classical Histories}
\label{classhist-gen}
The universal asymptotic expansion of the action \eqref{actexpn} supplies some information about the predictions for real, classical histories from a semiclassical wave function. Clearly the semiclassical wave function predicts no classical evolution for boundary metrics that have real values of $b$, because the {\gf dominant terms in the} action {\gf are} real and the classicality condition relating real and imaginary parts of the action cannot be satisfied. In the complex $\tau$-plane those geometries lie on the real axis.

However, it follows from the asymptotic solutions \eqref{expansions} that there are asymptotically horizontal curves in the complex $\tau$-plane that are displaced from the real axis along which the scale factor is purely imaginary and the scalar field real, resulting in a real boundary configuration with a complex asymptotic action \eqref{actexpn}. To identify these curves we write $u = e^{-\tau/l} = e^{-(x+iy)/l}$ and consider a horizontal curve in the $u$-plane defined by a constant value of $y=y_r$. To leading order in $u$ we have from \eqref{expansions} 
\be
a(u) = \frac{c}{u}= |c|e^{i\theta_c} e^{x/l}e^{iy/l}, \qquad \phi(u) = \alpha u^{\lambda_{-}^{\ }} = |\alpha|e^{i\theta_{\alpha}}e^{-\lambda_{-}^{\ }x/l}e^{-i\lambda_{-}^{\ }y/l}
\label{lead_phase}
\ee
where $c$ and $\alpha$ are the complex constants in \eqref{expansions} that are not determined by the asymptotic equations.
By tuning the phases $\theta_c$ and $\theta_{\alpha}$ so that
\be
\label{phases}
l \theta_c=-y_r, \qquad l \theta_{\alpha}=\lambda_{-}^{\ }(y_r+\pi/2).
\ee
we obtain an asymptotically real scalar profile and the following real, asymptotic metric along the $y=y_r +\pi/2$ curve,
\begin{equation}
ds^2 = dx^2 - |c|^2 e^{2x/l}\hijt dx^i dx^j
\label{localdS}
\end{equation}

{\gf The leading `counterterms' \eqref{ct} evaluated on these solutions are purely imaginary\footnote{This need not be true for scalar field models with masses below the BF bound for which the scalar counterterm has a real part.}. This means the classicality conditions hold for these boundary configurations {\hf because the real parts of the action are negligible compared to the rapidly varying imaginary parts.} Hence the wave function predicts the corresponding histories evolve classically.} The asymptotic solution \eqref{localdS} describes a classical history with an asymptotic locally de Sitter expansion \cite{Starobinsky83}. Thus a semiclassical wave function defined in terms of a negative $\Lambda$ theory predicts classical histories in which the expansion is driven by an effective positive cosmological constant $-\Lambda$.

We emphasize, however, that it does not follow from the asymptotic analysis alone that {\gf a given wave function has saddle points that asymptotically yield solutions of the form} \eqref{localdS} with \eqref{phases}. The asymptotic analysis only shows it is in principle possible for a wave function obeying the WDWE to predict such classical histories. Whether or not {\gf this is actually the case in a given theory} depends on the specific configuration and on the theory of the quantum state. By extension, the relative probabilities of different classical histories in the ensemble will depend on the choice of wave function, including its behavior for small scale factor.
In the next section we illustrate this for a particular solution of the WDW equation, namely the semiclassical no-boundary wave function.

\section{Example: The Holographic No-Boundary Wave Function}
\label{NBWF}

\subsection{The No-Boundary State for negative $\Lambda$}

The NBWF is usually formulated as a gravitational path integral involving the Euclidean action of  a gravitational theory with a positive cosmological constant and a positive matter potential.  The semiclassical predictions which are the focus of this paper follow from the saddle points determining the NBWF's semiclassical form. We have shown \cite{HH11} that the semiclassical NBWF can also be viewed as a wave function in a theory with a negative cosmological constant and a negative scalar potential.  This is potentially a more natural and useful formulation in fundamental physics because it raises the possibility that AdS/CFT can be used to express the semiclassical NBWF more precisely in terms of the partition functions of dual field theories on the (conformal) boundary of the four-disk.  Further, it is a promising starting point for defining the NBWF beyond the semiclassical approximation.

Specifically, the NBWF {\ce in its holographic form} is given by 
\be\label{dualNBWF}
\Psi [b,\tilde h_{ij}, \chi]=\frac{1}{Z_{\epsilon}[\tilde h_{ij},\tilde \chi]}\exp(-S_{ct}[b,\tilde h,\chi]/\hbar)
\ee
where $\tilde h_{ij}$ is the real boundary conformal structure which we normalize to have unit volume and $\epsilon \sim l/|b|$ is a UV cutoff. The source $\tilde \chi$ in the partition function $Z$ is the rescaled boundary value $b^{\lambda_{-}}\chi$. 
The $\exp(-S_{ct}/\hbar)$ factor in \eqref{dualNBWF} is given by \eqref{ct} and represents the universal part of the wave function discussed in Sections \ref{mss} and \ref{general}. As the subscript suggests, for positive {\cb signature}  boundary metrics $h$ this factor amounts to the counterterms often employed in AdS/CFT. For negative {\cb signature}  boundary metrics $h$, the $S_{ct}$ factor  has a large imaginary part and the integral curves of $S_{ct}$ describe asymptotically Lorentzian de Sitter histories if the classicality conditions on the wave function hold, as discussed above.

The $1/Z$ factor in \eqref{dualNBWF} governs the relative probabilities of different asymptotic configurations.
$Z$ is the partition function of a (deformed) Euclidean conformal field theory defined on the conformal boundary $\tilde h$. In a minisuperspace model involving gravity and a scalar field it is given by
\be\label{part}
Z[\tilde h_{ij},\tilde \chi] = \langle \exp \int d^3x \sqrt{\tilde h} \tilde \chi {\cal O} \rangle_{QFT}
\ee
with $\cal O$ the dual operator that couples to the source $\tilde \chi = b^{\lambda_{-}}\chi$ induced by the bulk scalar.
The brackets $\langle \cdots \rangle$ in \eqref{part} denote the functional integral average involving the boundary field theory action minimally coupled to the metric conformal structure $\tilde h_{ij}$. 

{\gf In field theory applications of AdS/CFT the value of $\tilde \chi$ is usually held fixed. However in a cosmological context one is interested in the wave function of the universe as a function of $\chi$. Indeed the dependence of $Z$ on the value of $\tilde \chi$ yields a {\hf probability} measure on different configurations on $\Sigma$ \cite{HH11}. Hence we are led to consider all deformations that correspond to real boundary configurations and for which the partition function converges. {\hf Considering all possible deformations is equivalent to considering all possible real domains in the asymptotic wave function. }

The Euclidean AdS/CFT correspondence \cite{Maldacena98} conjectures that, in an appropriate limit and {\cb to leading order in $\hbar$}, 
\be\label{duality}
Z[\tilde h_{ij},\tilde \chi] = \exp (-I^{reg}_{DW}[\tilde h_{ij},\tilde \chi]/\hbar) 
\ee
where $I^{reg}_{DW}$ is the `regularized' Euclidean AdS action\footnote{This is the Euclidean action plus the universal counterterms $S_{ct}$. The latter appear separately in the duality \eqref{dualNBWF}.} of a solution of the dual (super)gravity model which usually obeys the no-boundary condition of regularity in the interior and the boundary conditions $(\tilde h_{ij},\tilde \chi)$ asymptotically. In this paper we consider truncations of supergravity theories involving only scalar matter with a negative potential. The solutions that enter in \eqref{duality} are then AdS domain wall solutions of the Euclidean Einstein equations involving AdS gravity coupled to a scalar field in which the scalar has a nontrivial profile in the radial AdS direction and tends to the maximum of its potential near the boundary. 

In most applications of AdS/CFT the boundary is taken to be at infinity.
However in cosmology we are interested in the wave function of the universe at a large but finite value of the scale factor. The finite radius version of the duality is obtained from \eqref{duality} by integrating out a range of high-energy modes in $Z$, yielding a new partition function $Z_{\epsilon}$ where $\epsilon$ is a UV cutoff $\epsilon \sim l/|b|$.

\subsection{Two sets of saddle points}

The AdS/CFT correspondence provides a powerful way to evaluate the partition function \eqref{part} and thus the holographic NBWF \eqref{dualNBWF} in the limit where \eqref{duality} holds. We now illustrate this in the homogeneous isotropic minisuperspace {\cb model} spanned by the scale factor $b$ and the scalar field value $\chi$, in a toy model  {\cb of the form} \eqref{action}. {\cb We begin by calculating the saddle points that contribute to the leading order semiclassical approximation on the right hand side of \eqref{duality}.}

The homogeneous isotropic saddle points are of the form \eqref{4metric}. The field equations derived from \eqref{action} can be solved for $a(\lambda),\phi(\lambda)$ for any complex $N(\lambda)$ that is specified. Different choices of $N(\lambda)$ correspond to different contours in the complex $\tau$-plane. Contours start from the South Pole (SP) at $\lambda=\tau=0$ and end at the boundary $\lambda=1$ with $\tau(1)\equiv \upsilon$. Conversely, for any contour $\tau(\lambda)$ there is an $N(\lambda)\equiv d\tau(\lambda)/d\lambda$. Each contour connecting $\tau=0$ to $\tau=\upsilon$ is therefore a different representation of the same complex saddle point. The saddle point equations read
\begin{subequations}
\label{euceqns}
\begin{equation}
{\dot a}^2 -1-\frac{1}{l^2}a^2 +a^2\left(-{\dot \phi}^2 + 2 V(\phi)\right)=0,
\label{eucconstraint}
\end{equation}
\begin{equation}
\ddot\phi + 3\frac{\dot a}{a}\dot\phi  - \frac{d V}{d\phi}= 0 , 
\label{eucphieqn}
\end{equation}
\end{subequations}
where a dot denotes a derivative with respect to $\tau$. Solutions define functions $a(\tau)$ and $\phi(\tau)$ in the complex $\tau$-plane. A contour $C(0,\upsilon)$ representing a saddle point connects the SP at $\tau=0$ to a point $\upsilon$ where $a^2(\upsilon)$ and $\phi(\upsilon)$ take the real values $b^2$ and $\chi$ respectively. For any such contour the action can be expressed as
\be
I(b,\chi) = \frac{3\pi}{2}\int_{{C}(0,\upsilon)} d\tau a \left[ a^2\left(- \frac{1}{l^2}+ 2V(\phi)\right)-1\right].
\label{eucact_alt}
\ee

\begin{figure*}[t]
\includegraphics[width=3.0in]{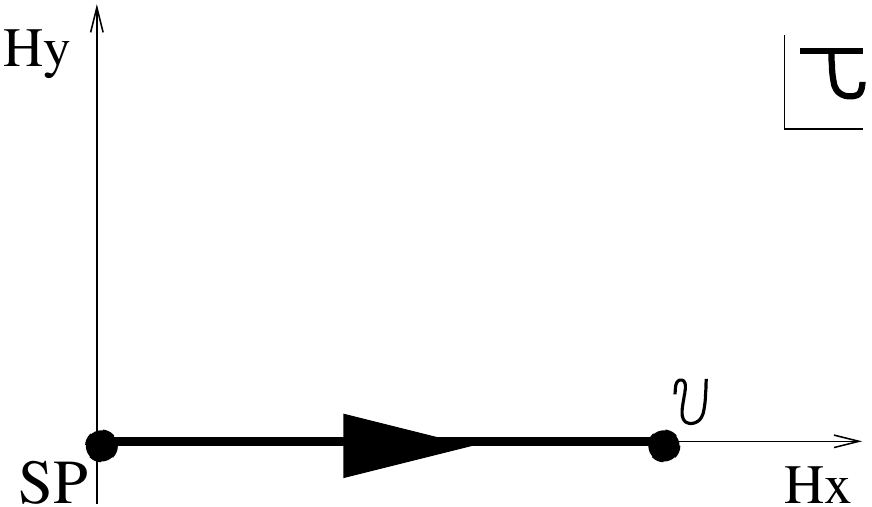} \hfill
\includegraphics[width=3.1in]{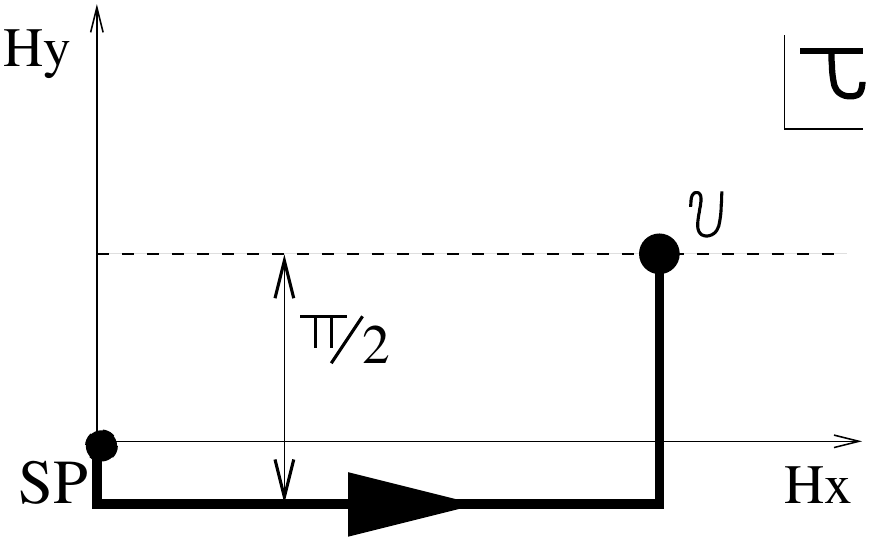}
\caption{{\it Left (a):} A representation in the complex $\tau$-plane of a real, Euclidean AdS domain wall saddle point, with $H \equiv 1/l$. {\it Right (b):} 
A domain wall with a complex scalar field profile along the horizontal branch of the contour for which the classicality conditions hold at the endpoint $\upsilon$.}
\label{contours}
 \end{figure*}

One set of saddle points can be found by starting with a real value of $\phi$ at the SP and integrating out along the real axis in the $\tau$-plane to an endpoint $\upsilon$ (see Fig \ref{contours}(a)). The scalar field is everywhere real along this contour and rolls towards the maximum of its negative potential. This yields a class of real Euclidean, asymptotically AdS, spherical domain walls of the type often considered in applications of AdS/CFT. 

The regularity conditions at the SP mean that the value of $\phi$ at the origin is the only free parameter there. 
Thus for fixed scale factor $b$ there is a one-parameter set of homogeneous saddle points of this kind that can be labeled either by the boundary value $\chi = \tilde \chi/b^{\lambda_{-}}$ or by the corresponding {\cb magnitude}  $\phi_0$ of the scalar field at the SP. The Euclidean action integral \eqref{eucact_alt} is purely real and approximately given by \cite{HHH08} 
\be
\label{sdpt1}
I_1(b,\gamma_{ij},\chi) = I^{reg}_{DW}-S_{ct}, \qquad  I^{reg}_{DW}(\chi) \approx - \frac{\pi}{4V(\phi(0))}
\ee
where the universal counterterms $S_{ct}$ are real and given by \eqref{ct}. Via AdS/CFT, the `regularized' action $I^{reg}_{DW}$ provides the saddle point approximation to the $1/Z$ factor in \eqref{dualNBWF}.

There is, however, a second class of saddle point solutions {\ce with real boundary data} that is particularly interesting from a cosmological viewpoint. These involve the same dual field theories but with complex scalar deformations $\tilde \chi$.

Starting with a complex value of $\phi$ at the SP and integrating out along the horizontal curve at $y=y_r$ in Fig \ref{contours}(b) yields an AdS domain wall solution with an approximately real scale factor -- since $\Lambda$ is real -- but with a complex scalar field profile\footnote{{\cf This clearly shows that the AdS/de Sitter connection derived here does not in general reduce to an analytic continuation. The complex saddle point solutions and their dual partition functions are genuinely different from the real ones except for massless scalar fields for which the asymptotic phase in the AdS regime vanishes \cite{HH11}.}} in the radial AdS direction\footnote{{\cf The value of $y_r$ depends on $\phi_0$ and goes to zero when $\phi_0 \rightarrow 0$.}}. The asymptotic expansions \eqref{expansions} show that if one tunes the phase of $\phi(0)$ at the SP so that it tends to $e^{i(\pi\ell/2)\lambda_{-}}$ along the $y=y_r$ curve, then the asymptotic three-metric and field are {\it both} real along a horizontal curve located $y=y_r +l \pi/2$ (indicated by the dotted line in Fig \ref{contours}(b)). There is one such curve for each value within a range of values of $\p0\equiv|\phi(0)|$ yielding a second set of homogeneous saddle points.

An interesting representation of the complex saddle points is based on the contour shown in Fig \ref{contours}(b). Along the horizontal part of this, the saddle point geometry is a complex version of the Euclidean AdS domain walls that made up the first set of saddle points. The {\cf last} vertical part of the contour in Fig \ref{contours}(b) corresponds to a transition region between the Euclidean AdS regime and the {\cf real boundary configuration}. The saddle point action \eqref{eucact_alt} is the sum of a contribution $I_h$ from the AdS domain wall and a contribution $I_v$ from the transition region to the endpoint $\upsilon$. The first contribution is given by
\be
\label{Ihor}
I_h = I^{reg}_{DW}(\tilde \chi) - S_{ct}(a,\tilde \chi), \qquad  I^{reg}_{DW} (\tilde \chi) \approx - \frac{\pi}{4V(\phi(0))}
\ee
where $a$ and $\tilde \chi$ {\ce are the values of the scale factor and scalar field {\crd at the point on  the contour  in Fig \ref{contours}(b) where it turns upwards}}. The {\cf gravitational} counterterms in \eqref{Ihor} are real because $a$ is real, but the regularized domain wall action $I^{reg}_{DW}$ is in general complex. The contribution from the vertical part of the contour to the saddle point action is universal and given by \cite{HH11}
\be
I_v = S_{ct}(a,\tilde \chi) - S_{ct}(b,\chi)
\ee
where $\chi=\tilde \chi/b^{\lambda_{-}}$. Using \eqref{Ihor} we get for the saddle point action
\be
\label{sdpt2}
I_2 (b,\gamma_{ij},\chi) = I^{reg}_{DW}(\tilde \chi) - S_{ct}(b,\chi)
\ee
where $b$ and hence $S_{ct}(b,\chi)$ are purely imaginary. This exhibits the universal asymptotic behavior for negative {\cb signature} boundary metrics discussed in Section \ref{general}. Using the AdS/CFT duality \eqref{duality} the regularized saddle point action can be used to evaluate the dual partition function with a complex deformation $\tilde \chi$ that defines the holographic NBWF.

{\cf The leading terms in \eqref{sdpt2} are purely imaginary and grow as $b^3$. This specifies an asymptotic de Sitter structure {\gf for which the classicality conditions hold} as discussed in Section \ref{general}. 
{\gf The NBWF for negative $\Lambda$ and $V$ thus predicts} an ensemble of classical, Lorentzian histories that are given by the integral curves of the imaginary part of the action of the regular complex saddle points. {\gf The histories were obtained explicitly in \cite{HHH08} for an everywhere quadratic potential, where it was found they} are asymptotically de Sitter with an effective cosmological constant $-\Lambda$, and with an early period of scalar field inflation driven by an effective positive potential $-V$ \cite{HHH08}. The amount of inflation is determined by $\phi_0$. The relative probabilities of different histories are given by the absolute value of $1/Z$ which follows in \eqref{sdpt2} using AdS/CFT. One recognizes {\cre in $1/Z$}  the familiar $-1/V(\phi_0)$ factor governing the no-boundary probabilities of histories with different amounts of inflation.}

\subsection{Amplitude for Classical Behavior}

{\cf The holographic NBWF admits two distinct classes of homogeneous {\hf isotropic} saddle points with real boundary data. These are essentially monotonic Euclidean AdS domain walls with either a real or a complex radial scalar field profile. Each class of saddle points describes a one-parameter set of real, four-dimensional histories. The histories can be labeled by the absolute value $\phi_0$ of the scalar field at the SP of the corresponding saddle point or, equivalently, by the boundary value $\chi$ at a given scale factor $b$. For a given boundary geometry and scalar field there is one saddle point in each class, with opposite signature of the boundary metric.

The quantum mechanical histories associated with both types of saddle points are significantly different.
The wave function does not predict the first class of histories {\cre to}  behave classically, since the action of the corresponding saddle points is purely real. By contrast, the action of the second class of saddle points has a large phase factor {\gf which, in the $V=(1/2)m^2\phi^2$ models we consider}, ensures the classicality conditions are satisfied for a range of $\phi_0$, leading to the prediction of an ensemble of histories that behave classically at least in the large volume regime where the semiclassical approximation to the wave function holds.

However a further question is whether the {\cb amplitudes for classical behavior are large in the NBWF.} Assuming that histories corresponding to different saddle points decohere, this can be seen from a comparison of the actions \eqref{sdpt1} and \eqref{sdpt2}. The real part of the Euclidean AdS action of the complex saddle points tends to a constant asymptotically whereas it grows as $-b^3$ for the real saddle points. Hence the complex saddle points corresponding to classical histories provide the overwhelmingly dominant contribution to the NBWF.

{\cd On the other hand} quantum states for which classicality has a low probability are {\cd not} ruled out. This is because our observations are necessarily conditioned on classical spacetime in our neighborhood so we have no direct test of the probability of classicality itself\footnote{For a recent discussion of conditional probabilities for observations in quantum cosmology see \cite{HHH10b}.}. Nevertheless it is an attractive feature of the NBWF that it {\cb gives large amplitudes for classicality.}}

\section{Summary}

We live in a quantum universe. Our observations of it on large distance scales are related to its classical behavior. Predictions of this classical behavior from a fundamental quantum mechanical theory of the universe's dynamics and quantum state are therefore of central importance. We have argued that accelerated expansion can be a prediction {\cd at low energies} of dynamical theories with a negative cosmological constant that specify wave functions satisfying the constraints of general relativity. The key ingredients in reaching this conclusion are as follows:

\begin{itemize}

\item A quantum mechanical definition of classical histories as ones with high probabilities for correlations in time governed by classical dynamical laws --- the classical Einstein equations in particular. 

\item A quantum state of the universe whose semiclassical (leading order in $\hbar$) approximation determines the ensemble of possible classical histories as defined above together with their probabilities.

\item A universal, large volume, complex, asymptotic semiclassical structure following from the WDWE that is exhibited by any wave function {\cd in the presence of a non-zero cosmological constant}. This complex structure can be described either by the action that supplies the semiclassical approximation to the wave function or by an equivalent asymptotic expansion of complex solutions to the Einstein equations. 

\item A prescription for extracting predictions for classical histories from the wave function's complex semiclassical structure that does not distinguish between different real domains but includes the classical histories arising from all of them in the classical ensemble. 

\end{itemize}

Given this {\ce general framework}, the argument proceeds as follows: The universal semiclassical asymptotic wave functions in theories with a negative cosmological constant describe two classes of real asymptotic histories --- asymptotically Euclidean AdS for boundary metrics with one signature and Lorentzian de Sitter for metrics with the opposite signature. {\cq Assuming boundaries with spherical topology} the classicality condition {\ce can be} satisfied only for the asymptotically de Sitter histories. Therefore negative $\Lambda$ theories can be consistent with our observations of classical accelerated expansion.  

The probabilities of different classical histories with accelerated expansion depend on the specific wave function assumed. As an illustration we calculated the probabilities for the case of the semiclassical no-boundary wave function, {\ce formally} defined in terms of the partition functions of (deformed) Euclidean CFTs on the (conformal) boundary of the four-disk, assuming a simple scalar matter model and AdS gravity in the bulk.  

This naturally raises the question for what class of dynamical theories can our results be expected to hold. We now turn to that.

 \section{Prospects for String Cosmology}

Treating the classical behavior of our universe as an emergent property in a fully quantum mechanical framework of cosmology opens up new possibilities for building models of inflation in string theory.
 
A familiar construction of string theory models of inflation involves `uplifting'  an AdS vacuum to a metastable  vacuum with a positive cosmological constant (see e.g. \cite{Kachru03,Silverstein11}). This paper shows that there is a wide class of alternative models in which classical accelerated expansion is understood as a low-energy prediction of the wave function of the universe in a fundamental theory with a negative cosmological constant. {\cq Given that string theory appears to be more secure with AdS boundary conditions this is an appealing prospect from a theoretical viewpoint.}

Toy models to which our analysis applies are supplied by the consistent truncations of the low energy limit of M theory compactified on $S^7$ involving only AdS gravity and one or several scalars with negative potential (see e.g. \cite{Hertog05}). The scalar potentials in these models satisfy the BF bound near the negative maximum, and fall off exponentially  at large values of the field. They act as positive, inflationary effective potentials in the classical `cosmological' domain of the complex asymptotic solutions that define the semiclassical approximation to the wave function. An ensemble of classical histories with epochs of scalar field driven inflation that asymptote to stable de Sitter space is predicted. The models used in this paper are examples of this. 

However, the inflationary histories {\cq in these consistent truncations} are unstable when viewed as low-energy solutions in the full theory\footnote{We thank Eva Silverstein for discussions of this point.}. This is because in the complete theory there are light scalars with a positive potential around the AdS vacuum. These give rise to instabilities in the cosmological regime where their potential has negative mass squared directions. It appears that a (meta)stable cosmology should be based on an AdS compactification in which all light scalar fields have zero or negative squared mass\footnote{It does not seem necessary to restrict to compactifications in which all scalars satisfy the BF bound, because only the Euclidean AdS theory enters in the wave function.}. In our framework, vacua that do not satisfy this condition are ruled out by our classical observations of a long-lived, accelerating universe. The requirement of a stable cosmology thus acts as a (strong) vacuum selection principle.

A qualitatively different constraint on the scalar potential comes from the classicality conditions \cite{HHH08}. A semiclassical wave function \eqref{semiclass} predicts classical evolution of space-time and matter fields in regions of superspace where its phase $S$ varies rapidly compared to the real part of the Euclidean action $I_R$.
The asymptotic structure implies that the classicality conditions hold for the empty de Sitter history, which is thus a universal prediction of wave functions in theories with a non-zero cosmological constant. In the presence of matter, however, the classicality conditions are not automatically satisfied even in the asymptotically de Sitter regime. Instead whether classical evolution emerges in this more general context depends both on the boundary configuration and on the specific choice of wave function. 

This is illustrated by the example of the NBWF with a quadratic scalar potential that we worked out in \cite{HHH08}. 
{\cq The results of that paper imply} that when the scalar mass is below the BF bound, a minimum amount of matter is required in order for the wave function in the saddle point approximation to predict classical cosmological evolution.
In particular there are no homogeneous isotropic histories for a range of scalar field values at the South Pole of the saddle point in a neighborhood of the AdS vacuum. This can be understood qualitatively from the form of the counterterms \eqref{ct}. For scalar masses below the BF bound the scalar counterterm is no longer purely imaginary. Since this contributes to the asymptotic action at leading order in the scale factor, the classicality {\cq constraints} will depend on the details of the configurations and the state. More generally, we expect that classical histories will emerge as a prediction of the NBWF only in theories where the scalar potential has sufficiently flat patches.

Predictions of classical histories depend both on the AdS compactification and on the quantum state of the universe. 
We have used the NBWF as a model of the quantum state but 
it would also be interesting to construct {\cq theories of cosmology} based on wave functions other than the NBWF. The universality of the asymptotic form of solutions to the WDWE means the AdS/de Sitter connection applies to other theories of the quantum state as well. In many situations, Euclidean AdS/CFT implements the no-boundary condition of regularity in the interior of the bulk. It therefore calculates amplitudes in the no-boundary state and this is encoded in the dual field theory. However, given that certain classical singularities can be resolved in string theory it is conceivable one can find more intricate models based on a different boundary condition that yield viable cosmologies. The holographic formulation of wave functions provided by Euclidean AdS/CFT furthermore has the appealing feature of unifying the theory of the state and the dynamics in a single entity --- the boundary field theory partition function.

It is in ways like this that quantum cosmology enables us not just to calculate quantum {\cq processes} near the big bang but also gives us a deeper understanding {\cq of} our universe at the classical level today.

\vskip .2in

\noindent{\bf Acknowledgments:} We thank Gary Horowitz, Alex Maloney, Don Marolf, Joe Polchinski and Eva Silverstein for helpful and stimulating discussions. Thanks are due to Lubos Motl and blogger HB for pointing out a typo in an early version. TH thanks the KITP and the Physics Department at UCSB for their hospitality during the completion of this work. JH and SWH thank Marc Henneaux and the International Solvay Institutes for their hospitality. The work of JH was supported in part by the US NSF grant PHY08-55415. The work of TH was supported in part by the FWO-Vlaanderen under the Odysseus program.

\end{document}